\documentclass[12pt]{article}

\usepackage{graphics}
\usepackage{cite}

\newcommand{\be}{\begin{equation}}
\newcommand{\ee}{\end{equation}}
\newcommand{\ba}{\begin{eqnarray}}
\newcommand{\ea}{\end{eqnarray}}
\newcommand{\half}{\ensuremath{\frac{1}{2}}}
\newcommand{\dt}{\frac{d}{dt}}
\newcommand{\neut}[1]{\ensuremath{\nu_{#1}}}
\newcommand{\aneut}[1]{\ensuremath{\bar{\nu}_{#1}}}
\newcommand{\mbf}[1]{\mathbf{#1}}

\newcommand{\ev}{\ensuremath{\; \mbox{eV}}}

\newcommand{\mev}{\ensuremath{\; \mbox{MeV}}}

\newcommand{\nn}{\nonumber \\*}
\newcommand{\ra}{\rightarrow}

\newcommand{\etal}{\emph{et.al.}}
\newcommand{\sgL}{{\it sign}($L$) }

\def\lsim{\;\raise0.3ex\hbox{$<$\kern-0.75em \raise-1.1ex\hbox{$\sim$}}\;}
\def\gsim{\;\raise0.3ex\hbox{$>$\kern-0.75em \raise-1.1ex\hbox{$\sim$}}\;}

\newcommand{\app}[3] {{\it  Astropart.\ Phys.\ }{{\bf #1} {(#2)} {#3}}}

\newcommand{\np}[3]  {{\it  Nucl.\ Phys.\ }{{\bf #1} {(#2)} {#3}}}
\newcommand{\pr}[3]  {{\it  Phys.\ Rev.\ }{{\bf #1} {(#2)} {#3}}}

\newcommand{\prl}[3] {{\it  Phys.\ Rev.\ Lett.\ }{{\bf #1} {(#2)} {#3}}}
\newcommand{\pl}[3]  {{\it  Phys.\ Lett.\ }{{\bf #1} {(#2)} {#3}}}

\begin{document}

%
%

\begin{titlepage}
\pagestyle{empty}
\baselineskip=21pt
%
\rightline{November 15th, 1999}
\vspace{.6in}

\begin{center} {\Large{\bf Physical origin of ``chaoticity'' \\
                           of neutrino asymmetry}}

\end{center}
\vspace{.3in}

\begin{center}

Antti Sorri \\

\vspace{.2in}

{\it  Physics Department, University of Helsinki, \\
          P.O.\ Box 9, FIN-00014 University of Helsinki }\\
\vspace{ .1in}

\end{center}

\vspace{0.3in}

\centerline{ {\bf Abstract} }
\baselineskip=18pt
\vspace{0.5cm} \noindent

We consider the indeterminacy in the sign of the neutrino asymmetry
generated by active-sterile neutrino oscillations in the early universe.
The dynamics of asymmetry growth is discussed in detail and 
the indeterminacy in the final sign of the asymmetry is shown to be
a real physical phenomenon. 
Recently published contradicting results are carefully considered
and the underlying assumptions leading to the disagreement are resolved.

\end{titlepage}

\baselineskip=20pt

%
%

The recent observation of strong zenith angle dependence of the atmospheric 
neutrino deficit by the Super-Kamiokande neutrino experiment has 
provided strong 
evidence for neutrino oscillations $\neut{\mu} - \neut{X}$,
where $\neut{X}$ is either $\neut{\tau}$ or a new, sterile neutrino 
$\neut{s}$ \cite{SK}. While the $\neut{\mu} - \neut{s}$
solution presently is less favored by SK data \cite{SK_s}, 
reconciling the existing data from all the neutrino experiments, including
LSND, is not possible unless there exists at least one sterile neutrino 
mixing with the active neutrinos.  Such mixing would have interesting 
consequences 
for primordial nucleosynthesis \cite{EKM,EKT,fschram,bdol,fvL,fvA,fvB} 
and CMB radiation \cite{cmb}.
For example, sterile neutrinos  could be brought into equilibrium prior 
to nucleosynthesis, increasing the energy density of the universe and 
thereby neutron-to-proton freeze out temperature, 
leading to more helium-4 being produced. This scenario has been numerically 
studied and strong limits to neutrino mixing parameters have been 
obtained \cite{EKT,fschram}.

Under certain conditions
active-sterile neutrino oscillations may also lead to exponential growth 
of neutrino asymmetry, as was first discovered by
Barbieri and Dolgov \cite{bdol}. Later Foot and Volkas observed 
that by this mechanism very 
large asymmetries could be generated, which would have a
significant  effect on the primordial nucleosynthesis \cite{fvL,fvA,fvB} 
by directly modifying directly the $n \leftrightarrow p$ reactions. 
Moreover, they showed that an asymmetry generated by $\neut{\tau} - \neut{s}$
mixing could suppress 
sterile neutrino production in $\neut{\mu}-\neut{s}$-sector, 
loosening the bounds of \cite{EKT}, 
according to which the SK atmospheric deficit could not be explained by
$\neut{\mu}-\neut{s}$  mixing.

Later it was found that this asymmetry generation is
chaotic in the sense that determining the sign of the final asymmetry \sgL
does not simply follow from the initial conditions \cite{shi_ch}. 
This phenomenon was studied
in \cite{EKS}, and it was shown that the indeterminacy is
associated with a region of mixing parameters, where asymmetry is rapidly 
oscillating right after the resonance. As a consequence  the the amount of 
Helium-4 produced cannot be precisely determined in such a scenario \cite{EKS}.

In a recent paper \cite{DHPS} it was claimed, however, that
\sgL is completely determined by the initial asymmetry,
and moreover, that
there is only a slight growth of asymmetry after the resonance.    
In this article we clarify the origin of indeterminacy
in \sgL  and show that it is a real physical
phenomenon, not disturbed by numerical inaccuracy. Instead, we will argue  
that the disagreement arises due to
overly simplifying approximations used in \cite{DHPS}.
Finally, we will
point out to a likely cause leading to observed suppression of the 
final magnitude of the asymmetry in \cite{DHPS}. 

\vspace{7mm}

In the early universe the coherent evolution of the neutrino states is
interrupted by frequent decohering collisions. Therefore
the evolution of the system needs to be studied using the density
matrix formalism. We parameterize the reduced density matrices of 
the neutrino and anti-neutrino ensembles as
\be 
\rho_{\nu}      \equiv \half      P_0 (1 + {\bf      P}) \; , \qquad
\rho_{\bar \nu} \equiv \half \bar P_0 (1 + {\bf \bar P}),
\label{rho}
\ee 
where each matrix is assumed to be diagonal in momentum space, while
each momentum state has $2 \times 2$-mixing matrix structure in the 
flavour space.
Solving
the full momentum dependent kinetic equations for these density 
matrices numerically \cite{EKT,QKE} is a very complicated task and
all attempts published to date have used some approximations to
simplify the problem. 
Here we use momentum averaged approximation  
{\it i.e.} we set 
$\mbf{P}(p) \ra \mbf{P}\left( \langle p \rangle \right)$, with
$\langle p \rangle \simeq 3.15 T$. This approach has been 
found to give a very good 
approximation of the $\neut{s}$ equilibration \cite{EKT}, and it will
be sufficient for the purposes of this letter.
The coupled equations are then (in the case of 
$\neut{\tau} - \neut{s}$
oscillations, other cases can be obtained easily by simple redefinitions
which are found for example in \cite{EKT}):
\def\phm{\phantom{-}}
\ba
\dot {\bf P}
    &=& {\bf      V}\times{\bf      P}- (D + \dt \log P_0) {\bf      P}_T
        +(1 - P_z) \dt \log P_0 \mbf{\hat z}, \nn
\dot {\bf {\bar P}}
    &=& {\bf \bar V}\times{\bf \bar P}
        - ( \bar D+ \dt \log \bar P_0) {\bf \bar P}_T 
        +(1 -\bar P_z) \dt \log \bar P_0 \mbf{\hat z}, \nn
\dot {P_0} 
    &=& \left\langle \Gamma^{\phm}\hspace{-0.3cm} 
        (\neut{\tau}\aneut{\tau} \ra \alpha \bar{\alpha})  
        \right\rangle 
        \left( n_{eq}^2 - n_{\neut{\tau} } n_{\aneut{\tau} } \right), \nn
\dot {\bar{P}_0}
    &=& \left\langle \bar{\Gamma} (\neut{\tau}\aneut{\tau} 
        \ra \alpha \bar\alpha)  \right\rangle 
        \left(n_{eq}^2 - n_{\neut{\tau} } n_{\aneut{\tau} } \right), 
\label{one state}
\ea
where $\dot x \equiv dx/dt$ and 
$\mbf{P}_T = P_x \mbf{\hat x} + P_y \mbf{\hat y}$. 
The damping 
coefficients for particles and anti-particles 
are $D \simeq \bar D \simeq 1.8G_F^2 T^5$ very accurately.
The rotation vector $\mbf{V}$ is
\be
  {\bf V} = V_x \;{\bf \hat{x}} + \bigl( V_0 + V_L \bigr) \; {\bf \hat{z}},
\ee
where
\ba
  V_x  &=&  \frac{\delta m^2}{2 \langle p \rangle} \sin 2 \theta ,  \nn
  V_0  &=&  \frac{\delta m^2}{2\langle p \rangle } 
            \cos 2 \theta + \delta V_{\tau}  , \nn
  V_L  &=&  -\sqrt{2} G_F N_{\gamma} \; L,
\label{Vcomp}
\ea
where $\theta$ is the vacuum mixing angle, 
$\delta m^2 = m^2_{\neut{s}} - m^2_{\neut{\tau}}$,
$N_{\gamma}$ is the photon number
density and the effective asymmetry $L$ in the potential $V_L$ is given by
\be
   L = - \half L_n + L_{\neut{e}} + L_{\neut{\mu}}
                    + 2 L_{\neut{\tau}}(P) 
     = \eta +  2 L_{\neut{\tau}}(P),
\label{alla}
\ee
in the case of an electrically neutral plasma. Asymmetry 
$L_{\neut{\tau}}$ is obtained from
\be
  L_{\neut{\tau}} = \frac{3}{8} \left( P_0(1+P_z) 
                    - \bar P_0 (1 + \bar P_z ) \right)
\label{asymm} 
\ee
and $L_n$ is the neutron asymmetry.
The potential term $\delta V_{\tau}$
is approximately \cite{EKT,notz}
\be
  \delta V_{\tau}
    =  17.8  G_F N_\gamma \frac{\langle p \rangle T}{2 M_Z^2}.
\ee
The rotation vector for anti-neutrinos is simply  
${\bf \bar V}(L) = {\bf V}(-L)$.

Neutrino and anti-neutrino ensembles  
are very strongly coupled in Eq.~(\ref{one state}) through the 
effective potential term $V_L(L)$, which
makes their numerical solution particularly difficult;
as long as neutrino asymmetry remains small,
there is a large cancellation in the Eq.~(\ref{asymm}),
leading to a potential loss of accuracy.
To overcome this problem we define new variables
\be
 P_{\alpha}^\pm \equiv P_{\alpha} \pm \bar{P}_{\alpha}.
\ee
In terms of these the Eq.~(\ref{one state})
for $\mbf{P^{\pm}}$  becomes
\begin{eqnarray}
\dot{P}_x^{\pm} & = &   -  V_0 P_y^{\pm}  - V_L P_y^{\mp} 
                        - \tilde D P_x^{\pm}, \nn
\dot{P}_y^{\pm} & = & \phm V_0 P_x^{\pm}  + V_L P_x^{\mp} 
                      -\tilde D P_y^{\pm} - V_x P_z^{\pm} ,\nn
\dot{P}_z^{\pm} & = & \phm V_x P_y^{\pm}  + A_{\pm} \left( 2 - P_z^+ \right) 
                      -A_{\mp} P_z^{\pm}, 
\label{neweqs}
\ea
where we have defined
\ba
A_+ &\equiv& \dt \log P^+_0 , \nn
A_- &=& \:\:\: 0            , \nn
\tilde D_{\phm} &=& D + A_+.
\ea
Finally, since for the averaged interaction rates 
$ \langle \Gamma \rangle = \langle \bar \Gamma \rangle $,
the difference $P_0^-$ is not affected by collisions and we find 
\ba
\dot P_0^+  &=&  2 \left\langle \Gamma \right\rangle 
                  \left( n_{eq}^2-n_{\neut{\tau}} n_{\aneut{\tau}} \right), \nn
\dot P_0^-  &=& 0.  
\label{p0}
\ea

\vspace{7mm}

Our objective is to study whether the sign of the 
final asymmetry, \sgL, follows deterministically from the initial conditions.
In \cite{shi_ch,EKS} it was found to be chaotic, 
whereas the authors of \cite{DHPS}
claim that \sgL is fully deterministic and equal to the sign of the initial
neutrino asymmetry. Similar results have been reported by other groups
as well \cite{fvA,fvB,fvL,dibari}. 

The key ingredient in the physics leading to
the growth of the asymmetry is the appearance of the resonance. Indeed,   
if the squared mass difference $\delta m^2 <0$, the effective potential
$V_0 \pm V_L$ goes through zero at the resonance temperature
\be
T_{c} \simeq 16.0 \; (|\delta m^2| \cos 2\theta)^{1/6} \; \mev,
\label{resT}
\ee
where $V_L \approx 0$ is assumed, as effective asymmetry $L$ is
driven to zero well before the resonance.
After the resonance the balance of the system abruptly changes 
which leads to a rapidly growing $L$. We have numerically solved
Eqs.~(\ref{neweqs}) and (\ref{p0}), and examples of the results are shown
in Fig.~\ref{pic:vardb}.
 
The behaviour of the system can be understood by a simple analogy of 
a ball rolling down a valley. After the resonance temperature $T_c$ 
the originally stable valley at $L=0$ becomes a ridge line separating
two new, degenerate valleys corresponding to solutions of $V_0 \pm V_L =0$. 
The system may first oscillate from one valley to another passing over the
ridge line. However, 
because of friction (represented by damping terms) it will
eventually settle into one or the other of the new valleys. 
It is easy to imagine that when there are many 
oscillations, even a very small difference in initial conditions may 
grow to a large phase difference at the time of settling down. 
In Fig.~\ref{pic:vardb} these effects are demonstrated for two initial
values: $\eta=10^{-10}$ (solid line) and 
$\eta=2 \times 10^{-10}$ (dashed line) and oscillation parameters
$\delta m^2 = -10^{-2} \ev^2$ and $\sin^2 2 \theta = 10^{-7.5}$. For
these parameters the resonance occurs at temperature $T_c=7.41 \mev$.
\begin{figure}[ht]
  \vspace{5mm}
  \begin{center}
    \includegraphics{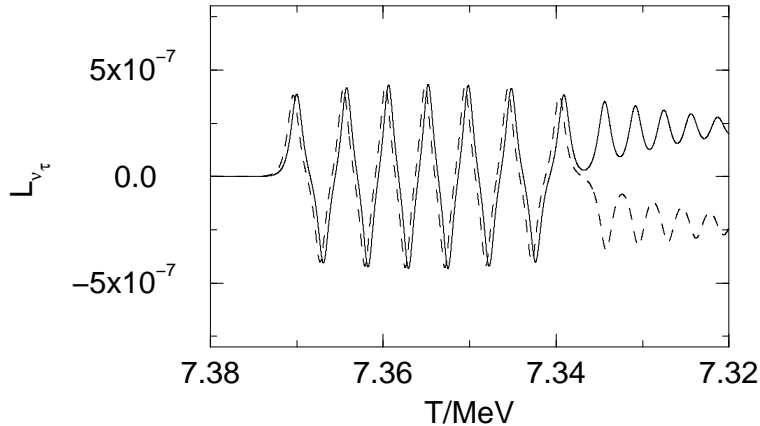}
  \end{center}
  \vspace{-5mm}
  \caption{The evolution of neutrino asymmetry at the resonance temperature
           for mixing parameters $\delta m^2 = -10^{-2} \ev^2$ and 
           $\sin^2 2 \theta = 10^{-7.5}$. Solid line corresponds to initial
           value $\eta= 10^{-10}$ and dashed line to 
           $\eta=2 \times 10^{-10}$.  }
  \label{pic:vardb}
\end{figure}

Let us point out that change of \sgL can be effected either by
variations in the oscillation parameters 
$\delta m^2$ and $\sin^2 2 \theta$
or variations in the initial asymmetry  
$\eta $ as was shown in \cite{EKS}.  
These two cases are physically quite different of course.
Varying oscillation parameters changes the shape of valleys
forming after the resonance. This is an important issue because there
will always be some experimental uncertainty in the measurements of masses
and mixings, leading to unavoidable uncertainty of SBBN predictions in
this scenario \cite{EKS}. Variations in $\eta$ are physical, 
for example due to local inhomogenieties in the baryon asymmetry created by
early phase transitions, and
correspond to deviations in the initial conditions (speed and direction 
of the ball) upon entering the resonance region.

In a recent article by Dolgov~\etal \cite{DHPS}, results 
were presented which are in sharp disagreement with ours. In particular, 
it was claimed that the sign of neutrino asymmetry is completely 
deterministic.  
Moreover, they  suggested that the chaotic behavior 
seen in \cite{EKS,shi_ch} would be due to accumulated errors in 
the numerical codes.  This suggestion is not correct. The numerical errors 
are completely under control in our computations. 
We have checked that allowing for the local error tolerance
magnitudes larger than actually employed to get our results, 
have no effects on them. Rather we will now argue that the disagreement
is due to ill-justified analytic approximations used in  \cite{DHPS}
to simplify the problem, which lead to artificial sign determinacy.

First approximation made in \cite{DHPS} was to neglect the term 
$V_x P_z^-$ in the equation for $P_y^-$ under the assumption that 
it is small in comparision to $V_L P_y^+$-term. However,
just before the resonance effective 
asymmetry $L$ is driven to zero, 
so that effective potential $V_L$ is in fact very small, 
and $V_x P_z^-$ should not be expected to be subdominant. 
We have indeed found that this term is of crucial importance for 
the initial asymmetry growth, as it prevents $L$ from getting
arbitrarily small value before the resonance, as will be discussed below.

Second approximation made in \cite{DHPS} has even 
more dramatic effects. There it was argued
that because Eq.~(\ref{neweqs})
for $P_x^-$ and $P_y^-$ can be scaled to a form 
$ \dot P_{x,y}^- = Q (a + b + c)$,
where $Q \approx 5.6 \times 10^4 \sqrt{|\cos 2\theta \delta m^2|}$ 
is a large parameter, it should be safe to set the derivatives 
$\dot P_{x,y}^-$ to zero. 
The rapid oscillations seen in our solutions indicate this approximation 
breaks down at the resonance temperature.

To study the effect of these approximations quantitatively,
we introduce them into  our equations. 
Dropping the term $V_x P_z^-$ and setting $\dot P^-_{x,y}$ to zero 
in the equations~(\ref{neweqs}) we then find the constraints 
\ba
  0 &=&   -  V_0 P_y^-  - V_L P_y^+ - \tilde D P_x^- ,  \nn
  0 &=& \phm V_0 P_x^-  + V_L P_x^+ - \tilde D P_y^-.     
  \label{pxyzero}
\ea
From these equations, one can solve the evolution of $P_x^-$ and 
$P_y^-$ algebraically with the result
\ba
  P_x^-  &=&   -  \frac{V_L}{V_0^2 + \tilde D^2} 
                  \left( \phm V_0 P_x^+ + \tilde D P_y^+ \right) \nn
  P_y^-  &=& \phm \frac{V_L}{V_0^2 + \tilde D^2} 
                  \left( -V_0 P_y^+ + \tilde D P_x^+ \right).
\label{offsol}
\ea
The remaining variables in Eqs.~(\ref{neweqs}) and (\ref{p0}) are 
then solved numerically. In Fig.~\ref{pic:comp} we plot the results
of a computation with (solid line) and without (dashed line) the 
implementation of the constraints~(\ref{offsol}) for same parameters
as in Fig.~\ref{pic:vardb}. The constrained solutions which fall on top of each
others in the figure are indeed fully deterministic and display 
no oscillation. This is to be
expected because the solutions~(\ref{offsol}) cannot produce sign 
changing oscillations in $L$, since when $L$ goes to zero so does $P_y^-$
and hence $\dot P_z^-$ is then strongly suppressed in the constrained case. 
This prohibits $L$ from changing sign, so that  \sgL is decided by
the initial value of $L$. When the evolution of the off-diagonal components
is neglected, the mechanism of the asymmetry growth is different, 
and instead of oscillating between the two new valleys,
the system slowly rolls down to one of them after the resonance. 
This is in fact 
common characteristic to so called static approximations \cite{fvA,DHPS}, 
where the differences between off-diagonals in the neutrino and antineutrino 
density matrices are ignored. 
\begin{figure}[ht]
  \vspace{5mm}
  \begin{center}
    \includegraphics{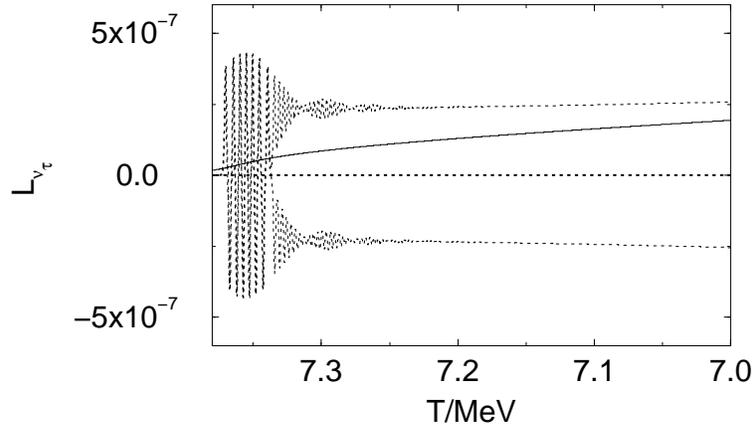}
  \end{center}
  \vspace{-5mm}
  \caption{Comparision of constrained (solid line) and 
           unconstrained (dotted line) solutions for the
           same parameters as in Fig.~\ref{pic:vardb}}
  \label{pic:comp}
\end{figure}

Let us now consider in detail how the complicated interplay
between the various terms in the Eq.~(\ref{neweqs}) leads to the 
initial exponential growth and oscillations of the asymmetry 
at the resonance temperature.    
Before the resonance $\mbf{P}^+$ components are practically independent
of $\mbf{P}^-$ components as effective asymmetry $L$ is very small.
However, off-diagonal components $P_x^+$ and $P_y^+$ do
grow to fairly large values. When $V_0$ changes sign at the resonance  
$P_x^+$ is rapidly driven to change sign, while the evolution of $P_y^+$ 
remains unaffected due to the additional term
$V_x P_z^+$. $P_z^+$ stays near its initial value until the off-diagonals 
begin to grow, after which it begins to diminish, 
signalling the sterile neutrino 
production. Overall the
evolution of $\mbf{P}^+$ components is smooth (see Fig.~\ref{pic:p+}), 
and the noticeable direct effect of the resonance is the changing of $P_x^+$.
\begin{figure}[ht]
  \vspace{5mm}
  \begin{center}
    \includegraphics{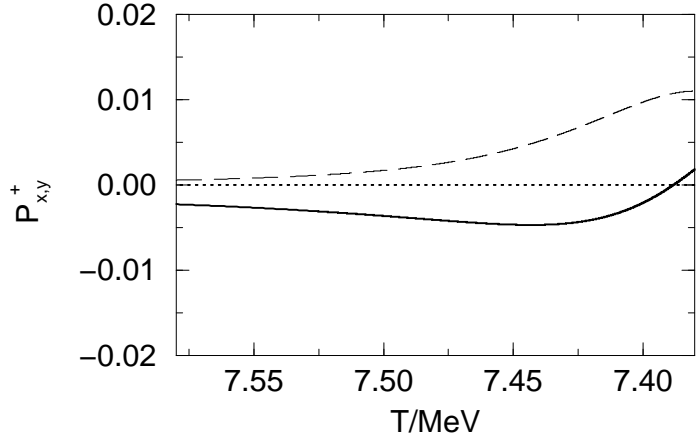}
  \end{center}
  \vspace{-5mm}
  \caption{Evolution of $P^+_x$ (solid line) and $P^+_y$ (dashed line)
           at the resonance.}
  \label{pic:p+}
\end{figure}

The evolution of $\mbf{P}^-$ components is more complicated.
Before the resonance neutrino and anti-neutrino ensembles
follow each other closely in the sense that off-diagonals
$P_{x,y}^-$ are small. The main force which is keeping 
the difference of neutrino and 
antineutrino off-diagonals small is the potential 
$V_L$ and not the effect of damping terms.
This can be confirmed by explicitly 
plotting individual terms appearing in the derivatives $\dot P_{x,y}^-$
(see Fig.~\ref{pic:termit}). The underlying mechanism driving
$\dot P_{x,y}^-$ close to zero before the resonance is then the 
cancellations between the
effective potential terms. The $V_x P_z^-$ term in Eq.~(\ref{neweqs})
prevents $L$ from stabilizing to an arbitrarily small value,
since when $L$ and off-diagonal components $P_{x,y}^-$ are driven towards zero
before the resonance, the small difference $P_z^-$ becomes important. 
The value of $V_L$ together with differences $P^-_{x,y}$  will cancel 
the effect of a non-zero $P_z^-$.  In this way the terms
$V_0 P_y^-$ and $V_L P_y^+$ cancel each other in the equation for $P_x^-$,
and $V_0 P_x^-$ together with $V_L P_x^+$ cancel $V_x P_z^-$ in the equation
for $P_y^-$. 
\begin{figure}[ht]
  \vspace{5mm}
  \begin{center}
    \includegraphics{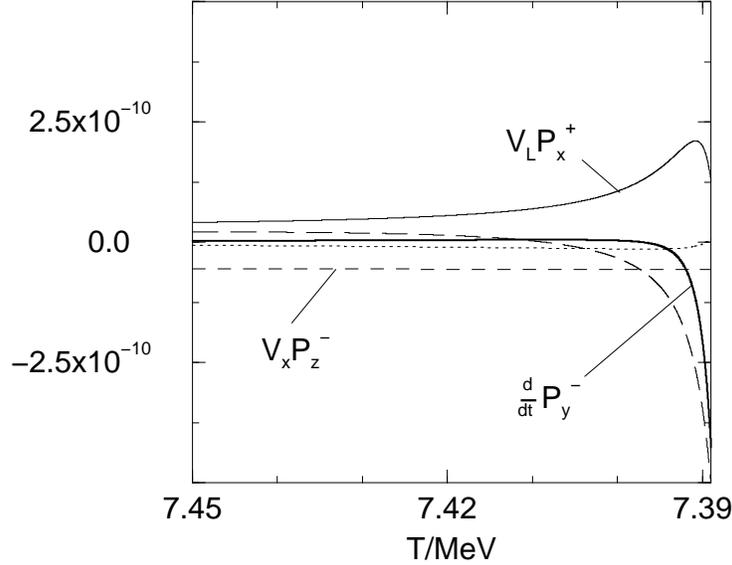}
  \end{center}
  \vspace{-5mm}
  \caption{The values of invidual terms in the Eq.~(\ref{neweqs}) for
           $P_y^-$ during the resonance: $\dot P_y^-$ (thick solid line),
           $V_L P_x^+$ (solid line), $V_0 P_x^-$ (long dashed line), 
           $V_x P_z^-$ (short dashed line) and $\tilde D P_y^-$ (dotted line)}
  \label{pic:termit}
\end{figure}

After $V_0$ changes sign the cancellation of the potentials in
$\dot P_x^-$ equation will not work, which turns around the effect
of $V_L$ in the $P_x^-$ equation. Moreover, as $P_x^+$ changes
sign due to the resonance, the effect of $V_L$ changes also in the 
$P_y^-$ equation. The difference between both off-diagonal components is now
growing due to a non-zero value of $L$ leading to a self supporting 
exponential growth of $L$.
We wish to stress that the magnitudes of all the effects discussed 
here are well above the 
numerical accuracy. It is true that before the resonance $L$ goes 
very close to zero, but this
is not relevant. What is relevant instead is that $P_z^-$ remains at 
a value of order $\eta$.   

It is interesting to note that the dynamics explained  above will
always lead to an initial growth of $L$ into the direction given by
the sign of $\eta$. This results agrees with the analytical considerations in 
\cite{fvB}. Nevertheless, as we have seen, this does not guarantee that
the final sign is that of $\eta$. 
One should bear in mind that the 
beginning of the asymmetry
growth is a very complicated phenomenon, where all the variables and
almost all the terms in the evolution equations are important 
to the outcome of 
the resonance and a complete account of the dynamics of the off-diagonal 
elements is of crucial importance. This makes it of course very difficult,
if not impossible, to find any sensible analytical approximation
to Eq.~(\ref{neweqs}).

There still remains one contradiction with respect to the results 
of  \cite{DHPS}
and the constrained solution obtained here; namely,
as seen from Fig.~\ref{pic:comp}, even the constrained solutions yield 
$L$ which grows to a large value. 
The quantitative study of this effect is beyond the validity of the method 
employed here. 
We may however point out a further approximation made in \cite{DHPS}, 
which appears to be the reason why they see a much weaker asymmetry growth.
The collision 
term in \cite{DHPS} for the active-active component, 
$\rho_{aa}$ of the density matrix~(\ref{rho}) is of the form
\be
 \Gamma(p) (\rho_{aa}(p) - f_{eq}),
 \label{coll}
\ee 
where in \cite{DHPS} $f_{eq}$ was taken to be the free Fermi 
distribution function $ f_{eq} = [1+\exp(p/T)]^{-1}$.
The authors in \cite{DHPS} note themselves that $f_{eq}$ should 
actually be the distribution function which includes a chemical potential,
$f_{eq}(\mu) = [1+\exp(p/T \mp \mu/T)]^{-1}$,
but they assume the  difference to be 
minor because the $\mu/T$ is small. However, one can show that  
$\mu/T \approx 0.7 L_{\neut{\tau}}$
and expand $f_{eq}(L_{\neut{\tau}})$ to give
\be
  f_{eq}(L_{\neut{\tau}}) \approx f_{eq}(0) \pm 0.7 L_{\neut{\tau}} 
                                  \frac{f_{eq}(0)}{1+\exp(p/T)}.       
\ee 
Approximating the second term of the expansion to zero in~(\ref{coll}) 
then means that the system is seeking the free Fermi
distribution instead of the correct equilibrium distribution which includes
an asymmetry. This gives rise to an artificial force  proportional 
to $\Gamma (p) L_{\neut{\tau}}$ resisting the growth of the asymmetry. 
Taking this into account, together with their previous assumptions leading to 
constraints~(\ref{offsol}), according to which the rate of asymmetry 
generation itself is weak
\be
 \dot P_z^- \propto P_y^- \propto V_L \propto L,
\ee
it appears likely that this force is able stop the asymmetry growth
before the system has reached the true bottom of the valley.
We chose not to pursue this issue further, since even the previous 
approximations has rendered the system unphysical by denying the 
possibility of chaoticity.

\vspace{7mm}

In this letter we have considered the indeterminacy or chaoticity in 
the sign of the neutrino asymmetry $L_{\neut{\tau}}$ arising from
$\neut{\tau}-\neut{s}$ mixing in the early universe. 
We carefully discussed the dynamics giving rise to the growth of asymmetry,
and unravelled the mechanism leading to the uncertaintity in 
\sgL. We confirmed that in the region of the parameter
space identified in \cite{EKS} the 
system is very sensitive to small variations in the initial conditions 
and mixing parameters. We have carefully checked that our numerical
methods are higly accurate, so that the effects of the variations
leading to a sign indeterminacy in the final asymmetry are physical.

As we pointed out, our results are in contradiction with the recent claims  
 \cite{DHPS}, according to which the sign of asymmetry
is completely deterministic and the asymmetry growth is small
compared to the results of \cite{fvA,fvB,fvL,shi_ch,EKS}. We have
showed that the results of \cite{DHPS} are in fact unphysical and that they
arise because of oversimplyfying approximations which artificially stabilize
the dynamics responsible for the chaoticity.

Indeed, we found that all the terms in the evolution equations and all the
components of the density matrix are important for
the dynamics of the system. Hence it appears unlikely that any simplifying
analytic approximation can be found that would describe the system adequately.
We have employed the momentum averaged equations,
as they are sufficient to study the chaoticity of the asymmetry growth
and to resolve the validity of approximations imposed in  \cite{DHPS}. 
Our unpublished results with a momentum dependent code support the results
of this letter, as well as do the results obtained by other groups working
with momentum dependent equations \cite{foot,dibcos}, although the
chaotic region appears to be somewhat smaller, when momentum dependence
is included. 

\vspace{7mm}

\textbf{Acknowledgements}

The author wishes to thank Kari Enqvist and Kimmo Kainulainen for enjoyable
collaboration on issues related to this letter, and in particular KK 
for suggesting this research and
his patient help in preparation of this manuscript. 
It is a pleasure to acknowledge the hospitality of 
NORDITA, Copenhagen, during the time this work was completed.

%
%

\end{document}